# Crystal structure built from a GeO$_6$-GeO$_5$ polyhedra network with high thermal stability: $\beta$–SrGe$_2$O$_5$

Christian A. Niedermeier,[1,*] Jun-ichi Yamaura,[2] Jiazhen Wu,[2] Xinyi He,[1] Takayoshi Katase,[1] Hideo Hosono[1,2] and Toshio Kamiya[1,2]

[1]Laboratory for Materials and Structures, Institute of Innovative Research, Tokyo Institute of Technology, 4259 Nagatsuta, Midori, Yokohama 226-8503, Japan

[2]Materials Research Center for Element Strategy, Tokyo Institute of Technology, 4259 Nagatsuta, Midori, Yokohama, 226-8503, Japan

*Corresponding author:* c-niedermeier@mces.titech.ac.jp

**Abstract:** By tackling the challenge of extending transparent oxide semiconductors to Ge-based oxides, we have found a not-yet-reported crystal structure, named $\beta$-SrGe$_2$O$_5$, which is composed of edge-sharing GeO$_6$ octahedra interconnected by GeO$_5$ bipyramids. Single crystals were successfully grown by the high-pressure flux method. $\beta$-SrGe$_2$O$_5$ has a band gap of 5.2 eV and a dispersive conduction band with an effective mass as small as 0.34 times the electron rest mass, which originates from the edge-sharing GeO$_6$ octahedra network. Although known compounds with octahedral GeO$_6$ coordination are commonly unstable at atmospheric pressure and elevated temperatures, $\beta$-SrGe$_2$O$_5$ exhibits thermal stability up to 700 °C.

Transparent conducting oxides (TCOs) and transparent oxide semiconductors (TOS) are emerging materials combining electrical conductivity and high optical transparency in the visible spectral region, which find important applications in optoelectronic devices. In particular, semiconductors with a band gap in the ultraviolet spectral region are increasingly gaining relevance for power electronics as they can tolerate high electric fields without device failure. Conventionally, TOS compounds contain post-transition metal cations such as $Zn^{2+}$, $In^{3+}$ and $Sn^{4+}$ of $(n-1)d^{10}ns^0$ electronic configurations ($n$ is the principal quantum number and should be $\geq 4$)[1,2]. Due to the wide spherical spread and overlap of the metal $n$s orbitals, an electron conduction path is formed[3], which results in a dispersive conduction band (CB) and small effective mass. Main constituents of TOS had been limited to the above heavy cations, and the extension of materials systems is a challenging task with great importance for further developing the current state of electronics technology.

In comparison with $Zn^{2+}$, $In^{3+}$ and $Sn^{4+}$, the effective ionic radius of $Ge^{4+}$ is rather small, and therefore, Ge-based TOS had not been known until the discovery of the large band dispersion and the high electrical conductivity in the cubic perovskite $SrGeO_3$ (space group $Pm\text{-}3m$)[4]. In its network of corner-sharing, regular $GeO_6$ octahedra, the $Ge^{4+}$ ions are 6-fold coordinated to $O^{2-}$ ions. This high symmetry configuration prohibits the hybridization of Ge 4s and O 2p orbitals at the $\Gamma$ point in the Brillouin zone, which forms the CB with exclusively Ge 4s non-bonding character and results in an exceptionally narrow 2.7 eV band gap and the first finding of high-density free electron doping in a Ge-based oxide. However, $Ge^{4+}$ typically occupies the 4-fold coordinated tetrahedral sites in the majority of Ge oxides that are thermodynamically stable at ambient conditions. On the other hand, compounds in which all $Ge^{4+}$ ions occupy 6-fold coordinated octahedral sites are rather exotic and commonly stable only in high pressure phases like the cubic perovskite $SrGeO_3$.

A recent high pressure study at 6 GPa reported α-$SrGe_2O_5$ (orthorhombic, $Cmca$), which crystal structure is built of stacked Ge oxide layers with corner-sharing $GeO_6$ octahedra and $GeO_4$ tetrahedra[5]. While the electrical properties and the electronic structure have not been reported, it may be expected that the separation of the Ge oxide layers in this structure and the large interatomic Ge-Ge distance due to the corner-sharing $GeO_4$-$GeO_6$ network impedes the electron transport. In contrast, the cubic perovskite $SrGeO_3$ is a very promising TCO, but the high synthesis pressure of 5 GPa[6] and the low thermal stability[7] are the key obstacles for practical applications. In particular, the crystalline-to-amorphous transition in cubic $SrGeO_3$ occurs at temperatures as low as 100 °C[8]. We thus target the synthesis of thermodynamically stable compounds built from a dense $GeO_6$ octahedral network, which would present very interesting novel candidates for high-mobility TOS.

In this paper, we report the discovery of a new crystal structure, named $\beta$-$SrGe_2O_5$ (orthorhombic, *Pnma*), with Ge ions in both 5-fold bipyramidal and 6-fold octahedral coordinations to O ions. This crystal structure is previously unknown and does not have any analogue in other material systems. According to density functional theory (DFT) calculations, $\beta$-$SrGe_2O_5$ is thermodynamically stable already at 1.4 GPa pressure, and the experiment confirms the good thermal stability up to 700 °C.

To find new Ge-based TOS, we inspected Ge-rich compounds of alkaline earth germanates as an analogue to the cubic perovskite $SrGeO_3$. Exploration of these materials for electronics is exciting due to two very advantageous features as compared to established TOS. First, the 1115 °C melting point of $GeO_2$ (trigonal, $P3_121$) is remarkably low as compared to the neighboring group IV oxides such as $SiO_2$ (1710 °C), $SnO_2$ (>2100 °C[9]) and $\beta$-$Ga_2O_3$ (1820 °C[10]). The low melting temperature makes single crystal growth from the melt significantly easier. Secondly, the alkaline earth cations in such ionic-covalent mixed bonding compounds work as network modifiers which facilitate the design of the $GeO_x$ polyhedra linkage and the Ge coordination.

Surprisingly, the ambient-pressure $CaO$-$GeO_2$ phase diagram[11] (Fig. S1(a)[12]) shows the stability of $CaGe_2O_5$ (monoclinic, *C2/c*), which is composed of a corner-connected $GeO_4$-$GeO_6$ network[13], whereas the compositional analogue of $SrGe_2O_5$ is not known[14] and solely the high-pressure $\alpha$-$SrGe_2O_5$ (orthorhombic, *Cmca*) phase has been synthesized at 6 GPa[5]. To find a dense crystal structure built from a network of $GeO_6$ octahedra, we investigated the Ge-rich part of the SrO-$GeO_2$ using differential thermal analysis (DTA). Remarkably, the compounds $SrGeO_3$ (monoclinic, *C2/c*) and $SrGe_4O_9$ (trigonal, *P321*) show a low 1164 °C eutectic temperature, while the $SrGe_2O_5$ phase is not observed (Fig. S1(b)).

We thus turned to investigate the stability of a dense $SrGe_2O_5$ phase at elevated pressure conditions. Applying the high-pressure synthesis and a SrCl$_2$/NaCl flux, we obtained colorless SrGe$_2$O$_5$ single crystals of up to 0.4 x 0.3 x 0.3 mm$^3$ size (Fig. 1(a)). We performed single crystal X-ray diffraction (XRD) measurement of the SrGe$_2$O$_5$ sample at room temperature. The initial structural model (orthorhombic, *Pmna*), named *β*-SrGe$_2$O$_5$, was suggested based on the unit cell dimensions and the reflection data. The final crystallographic parameters are listed in Table 1. The internal atomic coordinates, displacement parameters, and the displacement ellipsoid structure are given in Tables S1 and S2, and Fig. S2, respectively. The Rietveld refinement of the Bragg-Brentano powder XRD pattern of ground single crystals is well converged using the determined *β*-SrGe$_2$O$_5$ crystal structure (Fig. 1(b)), guaranteeing the phase stability after the grinding process.

Table 1: Crystallographic parameters for *β*-SrGe$_2$O$_5$.

| Chemical formula | SrGe$_2$O$_5$ |
|---|---|
| Crystal system | Orthorhombic |
| Space group | *Pnma* (#62) |
| Temperature (K) | 296 |
| Density, calc. (g/cm$^3$) | 5.65 |
| Unit cell volume (Å$^3$) | 368.001(15) |
| Lattice parameters *a*, *b*, *c* (Å) | 6.7530(2), 5.8725(1), 9.2796(2) |
| Calc. lattice parameters *a*, *b*, *c* (Å)[a] | 6.7303, 5.8748, 9.2990 |
| Crystal size (μm$^3$) | 70 x 50 x 50 |
| Radiation source | Mo K$_\alpha$ |
| Absorption correction | Numerical, NUMABS[15] |
| No. of measured reflections | 10264 |
| No. of independent reflections | 1398 |
| $R_{int}$[b] | 0.052 |
| (sin θ/λ)$_{max}$ (Å$^{-1}$) | 0.5 |
| $R_1$ [$F^2 > 2σ(F^2)$], $wR_2(F^2)$, $S$[b] | 0.049, 0.146, 1.13 |
| No. of parameters | 47 |
| Δρ$_{max}$, Δρ$_{max}$ (eÅ$^{-3}$) | 2.93, -2.55 |

[a]Obtained by hybrid DFT calculation using the PBE0 functional.
[b]See the SHELX manual[16] for definition of the residual and goodness of fit factors.

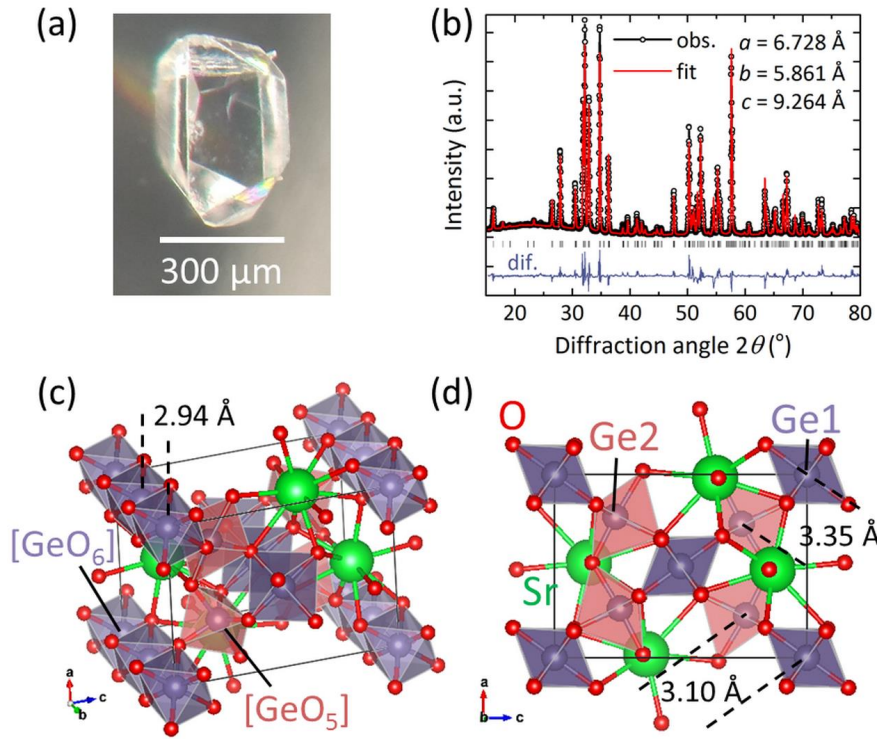

Fig. 1: (a) Optical micrograph of a transparent $\beta$-SrGe$_2$O$_5$ single crystal. (b) Bragg-Brentano powder XRD pattern of ground $\beta$-SrGe$_2$O$_5$ single crystals (black data points) with the result of Rietveld refinement (red line). (c,d) The $\beta$-SrGe$_2$O$_5$ (orthorhombic, *Pnma*) crystal structure with 5-fold and 6-fold coordinated Ge indicated by the red and purple polyhedra, respectively.

The crystal structure of $\beta$-SrGe$_2$O$_5$, which has an orthorhombic lattice and belongs to the Pnma space group, is illustrated in Figs. 1(c) and 1(d). The Sr$^{2+}$ ion is 10-fold coordinated to the surrounding O$^{2-}$ ions. The GeO$_6$ octahedra form edge-sharing one-dimensional chains along the b-axis. These GeO$_6$ chains are further corner-linked to GeO$_5$ triangular bipyramids, resulting in a three-dimensional cross-linked network with GeO$_5$-GeO$_6$ chains along the [1-11] and [11-1] crystallographic directions. The Ge-O bond lengths and O-Ge-O bond angles in the GeO$_6$ octahedron are 1.88–1.96 Å and 80.7–99.3°, indicating a minor distortion. The present phase shows some structural similarity to CaGe$_2$O$_5$ (orthorhombic, *Pbam*) obtained at 8 GPa pressure[17], in which Ge is 5- and 6-fold coordinated on square pyramidal and octahedral sites, respectively. In CaGe$_2$O$_5$, the degree of GeO$_6$ octahedral distortion is similar to that in $\beta$-SrGe$_2$O$_5$, showing Ge-O bond lengths and O-Ge-O bond angles of 1.87–1.94 Å and 82.4–97.3°, respectively. In $\alpha$-

SrGe$_2$O$_5$ (orthorhombic, *Cmca*), the Ge-O bond lengths are spread most widely from 1.84–1.99 Å. The detailed comparison of the Ge-O bond lengths and O-Ge-O bond angles in these structures, with reference to the high-symmetry cubic SrGeO$_3$ perovskite, is summarized in Fig. S3.

The interatomic Ge-Ge distance between the corner-linked GeO$_6$ octahedra in cubic SrGeO$_3$ of 3.80 Å is significantly larger than that of the edge-sharing GeO$_6$ octahedra in *β*-SrGe$_2$O$_5$ (2.94 Å) and CaGe$_2$O$_5$ (orthorhombic, *Pbam*) (2.82 Å). The Ge1-Ge2 interatomic distances within the corner-sharing GeO$_5$-GeO$_6$ polyhedral chains in *β*-SrGe$_2$O$_5$ are 3.10 Å and 3.35 Å (Fig. 1(d)), which are slightly longer than those between the edge-sharing GeO$_6$ octahedra. In CaGe$_2$O$_5$, the analogous Ge1-Ge2 distances are longer (3.36 Å and 3.43 Å), despite the smaller Ca$^{2+}$ effective ionic radius as compared to Sr$^{2+}$. In general, a small interatomic distance between the Ge4s orbitals, which form the CB, is beneficial for forming a high-mobility electron conduction path due to larger overlaps[3]. In the case of cubic SrGeO$_3$, however, the coordination symmetry is the key factor for determining the electron transport since the conduction path is formed by a Ge-O non-bonding state at the Γ point and Ge-O anti-bonding states at the Brillouin zone boundaries[18,19]. The Ge-O-Ge bond angle between two GeO$_6$ octahedra is exactly 180° which presents the largest possible interatomic distance of two Ge atoms in the network of GeO$_6$ octahedra.

Hybrid DFT calculations using the PBE0 functional[20] were performed to obtain the *β*-SrGe$_2$O$_5$ (orthorhombic, *Pnma*) ground state crystal structure. The lattice parameters are 6.73, 5.87 and 9.30 Å, which is within 0.4% relative error compared to the experimental values at room temperature (Table 1). The *β*-SrGe$_2$O$_5$ electronic band structure shows a 5.17 eV indirect band gap, with the CB minimum located at the Γ point and the valence band (VB) maximum located at the T [0 1/2 1/2] point of the Brillouin zone (Fig. 2(a)). The CB is dominated by Ge 4s and O 2p states, while the VB is mainly comprised of O 2p states (Fig. 2(b)). The *β*-SrGe$_2$O$_5$ CB is dispersive around the Γ point, with an effective mass as small as 0.34 times the electron rest mass $m_e$ in the Γ-Y direction (Table S3). The large dispersion along this direction is thus attributed to the chain of edge-sharing GeO$_6$ octahedra along the b-axis. The value of the effective mass is comparable to 0.28 $m_e$ for the prototype TCO SnO$_2$[21], which is also built of edge- and corner-sharing SnO$_6$ octahedra, similar to *β*-SrGe$_2$O$_5$.

We measured the reflection ellipsometry spectrum from 0.6–5.2 eV photon energy for a mirror-polished $\beta$-SrGe$_2$O$_5$ polycrystalline pellet. The steep increase in the real $\varepsilon_1$ and imaginary part $\varepsilon_2$ of the dielectric function approaching 5.2 eV energy is consistent with the calculated 5.20 eV direct electronic band gap (Fig. 2(c)). Using density functional perturbation theory[22], the frequency-dependent real and imaginary dielectric tensor elements $\varepsilon_x$, $\varepsilon_y$ and $\varepsilon_z$ are calculated and their averages are given for comparison.

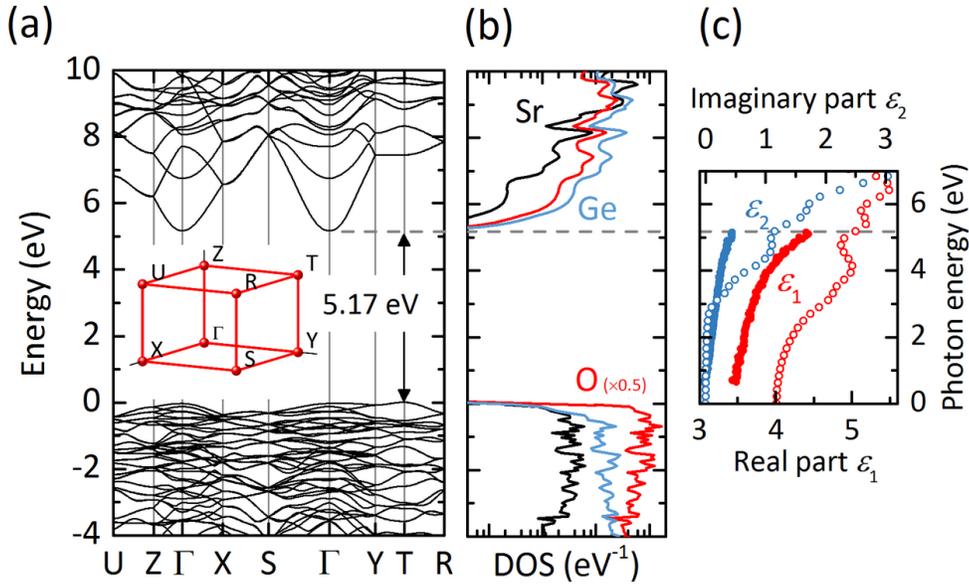

Fig. 2: (a) PBE0 band structure along high symmetry directions in the first Brillouin zone indicates a large CB dispersion and a 5.17 eV indirect band gap. (b) The density of states (DOS) in logarithmic scale shows that the CB is derived from Ge and O states while mainly O states contribute to the VB. (c) The measured real $\varepsilon_1$ and imaginary part $\varepsilon_2$ of the dielectric function of a mirror-polished $\beta$-SrGe$_2$O$_5$ polycrystalline pellet is given by the filled circles. For comparison, the averages of the calculated real and imaginary dielectric tensor elements $\varepsilon_x$, $\varepsilon_y$ and $\varepsilon_z$ are given by the open circles.

To assess the eligibility as TOS and investigate electrical properties, we doped Sb to substitute Ge in $\beta$-SrGe$_2$O$_5$ single crystals and $\beta$-Sr(Ge$_{0.95}$Sb$_{0.05}$)$_2$O$_5$ polycrystalline pellets. The Sb-doped $\beta$-SrGe$_2$O$_5$ single crystals show a dark blue color, but the polycrystalline pellets are electrically insulating and free carriers cannot be detected in the infrared reflectance spectrum (Fig. S4). The efficient impurity doping and underlying compensation mechanisms shall be the subject of further investigation.

Since compounds with octahedral GeO$_6$ coordination are commonly unstable in particular at atmospheric pressure, we investigate the β-SrGe$_2$O$_5$ thermodynamic stability theoretically using DFT calculations. The calculated formation enthalpy

$$\Delta H_f^{SrGe_2O_5} = \mu_{Sr} + 2\mu_{Ge} + 5\mu_O \qquad (1)$$

is constrained by the chemical potentials $\mu$ of elemental Sr, Ge and molecular O$_2$, with the origins at zero pressure and absolute zero temperature. The equilibrium conditions to prevent precipitation into solid elemental Sr (cubic, *Fm-3m*), Ge (cubic, *Fd-3m*) and molecular O$_2$, $\mu_{Sr} \leq 0$, $\mu_{Ge} \leq 0$ and $\mu_O \leq 0$ define the stable potential window boundaries. Considering all known phases in the quasi-binary SrO-GeO$_2$ system, including SrO (cubic, *Fm-3m*) and Sr$_2$GeO$_4$ (orthorhombic, *Pna2$_1$*), the most stable phases are determined within the given potential range. For zero external pressure, we confirm that the β-SrGe$_2$O$_5$ phase is not thermodynamically stable under equilibrium (Fig. 3(a)).

To investigate the β-SrGe$_2$O$_5$ stability at elevated pressures, the energy calculations of all (meta)stable compounds, including the high-pressure polymorphs in the SrO-GeO$_2$ system, were performed for a fixed set of volumes around the equilibrium volume (Fig. S5, Table S5). The Murnaghan equation[23] was used to derive an equation of state for fitting of the obtained energy-volume curves and the formation enthalpy is derived as a function of pressure[12]. Already at pressures as low as 2 GPa, a β-SrGe$_2$O$_5$ potential window is obtained with phase boundaries to SrGeO$_3$ (triclinic, *P-1*) and GeO$_2$ (tetragonal, *P4$_2$/mnm*) (Fig. 3(b)).

The transition pressure for the β-SrGe$_2$O$_5$ formation is determined by comparison with the enthalpy of the decomposition products, SrGeO$_3$ + GeO$_2$. The decomposition is energetically favored only for pressures below 1.4 GPa (Fig. 3(c)). Since the enthalpy difference between β-SrGe$_2$O$_5$ and the decomposition products at zero pressure is only about 0.01 eV/atom, it shall be possible to synthesize β-SrGe$_2$O$_5$ at ambient pressure under optimized growth conditions. For example, providing a crystal seed or epitaxial template favorable for the nucleation and growth of β-SrGe$_2$O$_5$ may realize such conditions. Moreover, the calculation shows that β-SrGe$_2$O$_5$ is the most stable modification at ambient pressure in comparison with the high-pressure α-SrGe$_2$O$_5$ (orthorhombic, *Cmca*) phase obtained at 6 GPa[5].

To investigate the thermal stability of $\beta$-SrGe$_2$O$_5$, the single crystals were ground to powder and annealed in air at ambient pressure. After thermal annealing of $\beta$-SrGe$_2$O$_5$ at 700 °C for 1 h, the XRD pattern shows that the phase remains stable and decomposition into SrGeO$_3$, SrGe$_4$O$_9$ and GeO$_2$ occurs only at elevated temperatures (Fig. 3(d)). The thermal stability facilitates growth of single crystals at ambient pressure through a suitable flux by adding a seed crystal when cooling from 700 °C temperature.

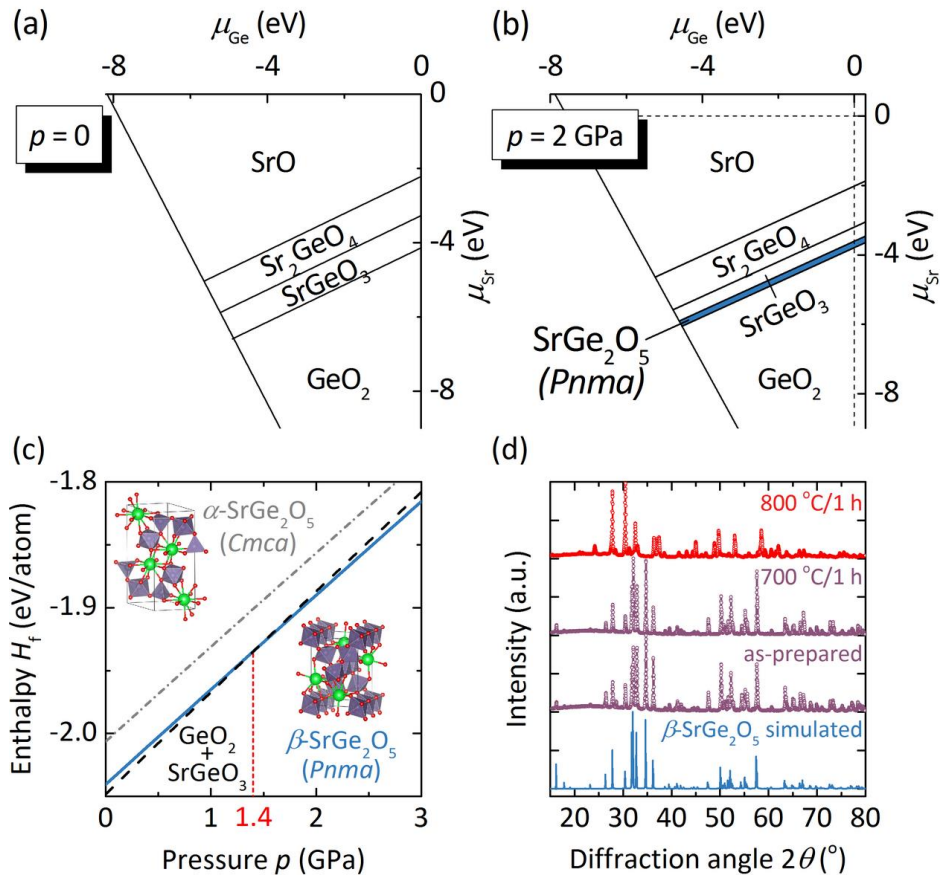

Fig. 3: (a) Absence of a chemical potential window for $\beta$-SrGe$_2$O$_5$ at zero pressure shows that the phase is not thermodynamically stable at the standard condition. (b) At a pressure as low as 2 GPa, $\beta$-SrGe$_2$O$_5$ becomes stable within the chemical potential window indicated by the blue area. (c) Formation of a phase-separated SrGeO$_3$-GeO$_2$ system is energetically favored over $\beta$-SrGe$_2$O$_5$ only for pressures below 1.4 GPa. (d) Bragg-Brentano XRD patterns of ground and annealed single crystals confirm the thermal stability of $\beta$-SrGe$_2$O$_5$ up to 700 °C temperature.

In summary, we have discovered a previously unknown crystal structure, orthorhombic $\beta$-SrGe$_2$O$_5$ which belongs to the *Pnma* space group. The single crystals were successfully grown using the high-pressure flux method. $\beta$-SrGe$_2$O$_5$ has a band gap of 5.2 eV and shows a large CB dispersion with an effective mass as small as 0.34 $m_e$ as a result of the edge-sharing GeO$_6$ octahedral network. The calculated pressure required for the $\beta$-SrGe$_2$O$_5$ synthesis is only about 1.4 GPa and the material shows good thermal stability at temperatures up to 700 °C.

## Supporting Information

CaO-GeO$_2$ and SrO-GeO$_2$ phase diagrams, $\beta$-SrGe$_2$O$_5$ crystallographic information, calculated effective mass along different symmetry directions, characterization of Sb-doped $\beta$-SrGe$_2$O$_5$, enthalpy equation of state for compounds in the SrO-GeO$_2$ system as a function of pressure, experimental and computational methods.

## Acknowledgements

The work at Tokyo Institute of Technology was supported by the MEXT Element Strategy Initiative to Form Core Research Center. C.A.N. acknowledges the support through a fellowship granted by the German Research Foundation (DFG) for proposal NI1834.

# Supporting information for

# Crystal structure built from a GeO$_6$-GeO$_5$ polyhedra network with high thermal stability: $\beta$–SrGe$_2$O$_5$

Christian A. Niedermeier,[1,*] Jun-ichi Yamaura,[2] Jiazhen Wu,[2] Xinyi He,[1] Takayoshi Katase,[1] Hideo Hosono[1,2] and Toshio Kamiya[1,2]

[1]Laboratory for Materials and Structures, Institute of Innovative Research, Tokyo Institute of Technology, 4259 Nagatsuta, Midori, Yokohama 226-8503, Japan

[2]Materials Research Center for Element Strategy, Tokyo Institute of Technology, 4259 Nagatsuta, Midori, Yokohama, 226-8503, Japan

*Corresponding author: c-niedermeier@mces.titech.ac.jp

## CaO–GeO₂ and SrO–GeO₂ phase diagrams

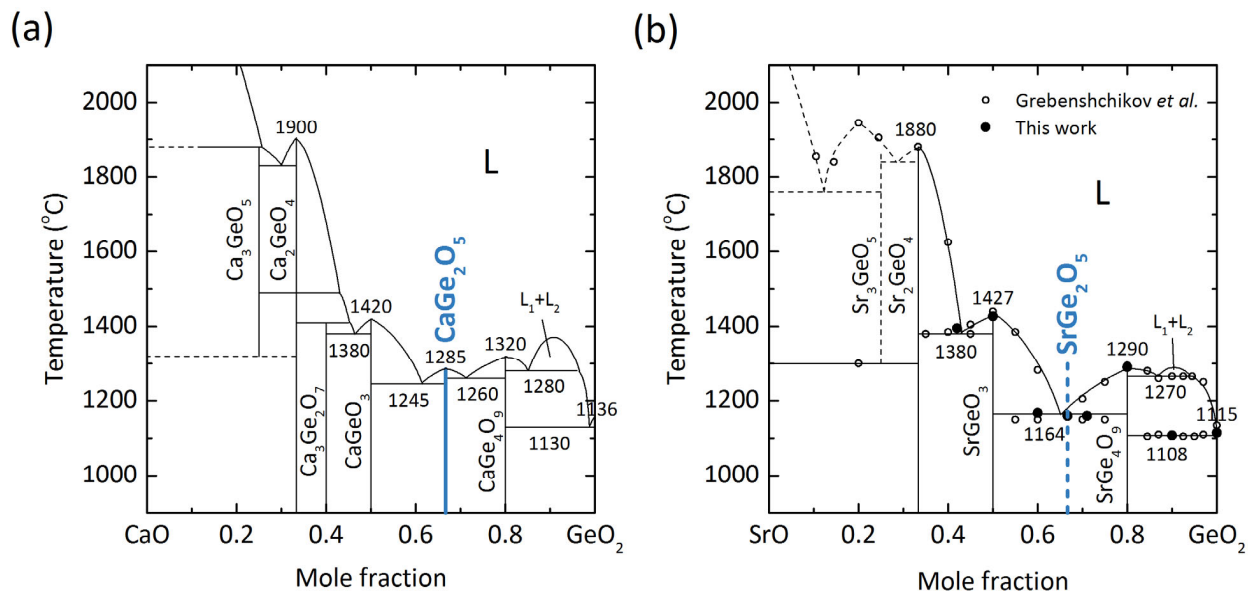

Fig. S1: (a) The CaO-GeO₂ phase diagram reprinted from Shirvinskaya et al.[1] indicates the existence of an ambient-pressure stable CaGe₂O₅ phase. (b) The SrO-GeO₂ phase diagram with the open data points is cited from Grebenshchikov et al.[2] In this reference, the Sr₃GeO₅ phase erroneously appears at a GeO₂ mole fraction of 0.2, which we corrected to 0.25 without experimental confirmation. The closed circles present our data obtained by DTA. The composition of SrGe₂O₅ is indicated with the blue dashed line.

# β-SrGe$_2$O$_5$ Crystallographic Information

Table S1: Internal atomic coordinates *x*, *y* and *z* and equivalent isotropic displacement parameters $U_{eq}$ (Å$^2$) for β-SrGe$_2$O$_5$.

| Atom | x | y | z | $U_{eq}$ |
|---|---|---|---|---|
| Sr | 0.9811(1) | 0.75 | 0.63611(6) | 0.0071(1) |
| Ge1 | 0.5 | 0.5 | 0.5 | 0.0040(1) |
| Ge2 | 0.7307(1) | 0.25 | 0.74124(7) | 0.0043(1) |
| O1 | 0.8430(7) | 0.25 | 0.9194(5) | 0.0057(6) |
| O2 | 0.6441(7) | 0.25 | 0.4244(5) | 0.0060(6) |
| O3 | 0.6630(5) | 0.5141(6) | 0.6635(3) | 0.0063(5) |
| O4 | 0.9812(8) | 0.25 | 0.6493(5) | 0.0100(8) |

Table S2: Atomic displacement parameters (ADP) in Å$^2$ for β-SrGe$_2$O$_5$.

| Atom | $U^{11}$ | $U^{22}$ | $U^{33}$ | $U^{12}$ | $U^{13}$ | $U^{23}$ |
|---|---|---|---|---|---|---|
| Sr | 0.0062(2) | 0.0095(2) | 0.0056(2) | 0 | 0.00021(15) | 0 |
| Ge1 | 0.0045(2) | 0.0047(2) | 0.0028(2) | 0.00013(17) | -0.00050(15) | 0.00015(17) |
| Ge2 | 0.0052(2) | 0.0049(2) | 0.0028(2) | 0 | -0.00107(16) | 0 |
| O1 | 0.0083(16) | 0.0075(16) | 0.0013(12) | 0 | -0.0020(11) | 0 |
| O2 | 0.0064(15) | 0.0068(16) | 0.0048(14) | 0 | 0.0004(12) | 0 |
| O3 | 0.0092(11) | 0.0045(11) | 0.0053(10) | 0.0010(9) | -0.0025(8) | 0.0002(9) |
| O4 | 0.0057(16) | 0.020(2) | 0.0047(15) | 0 | 0.0008(12) | 0 |

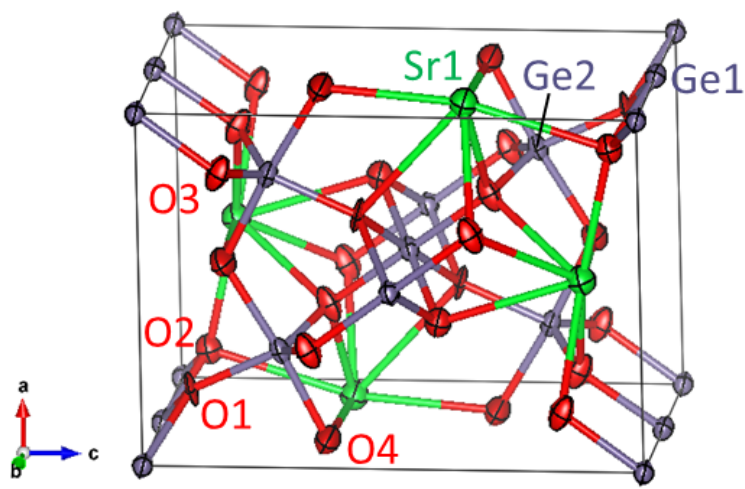

Fig. S2: β-SrGe$_2$O$_5$ crystal structure with internal atomic positions presented by displacement ellipsoids of 99% probability density as determined by the data in Tables S1 and S2.

Table S3: Comments on Level C alerts of the checkCIF/PLATON report for the structure refinement result of β-SrGe$_2$O$_5$.

| Alert | Problem | Comment |
| --- | --- | --- |
| DIFMX02 ALERT 1 C | The maximum difference density is > 0.1*ZMAX*0.75. The relevant atom site should be identified. | The maximum difference density occurs near the Ge atoms. |
| PLAT097 ALERT 2 C | Large reported max. (positive) residual density, 3.08 eA$^{-3}$ | The difference densities are located near the Sr and Ge atoms. These are ghost peaks which result from the difference Fourier synthesis. |
| PLAT213 ALERT 2 C | Atom O1 has ADP max/min ratio, 3.4 oblate | We checked its origin carefully, including the lowering of the symmetry. |
| PLAT250 ALERT 2 C | Large U3/U1 ratio for average U(i,j) tensor, 2.4 note | The O1 and O4 atoms indicate large ADP. |

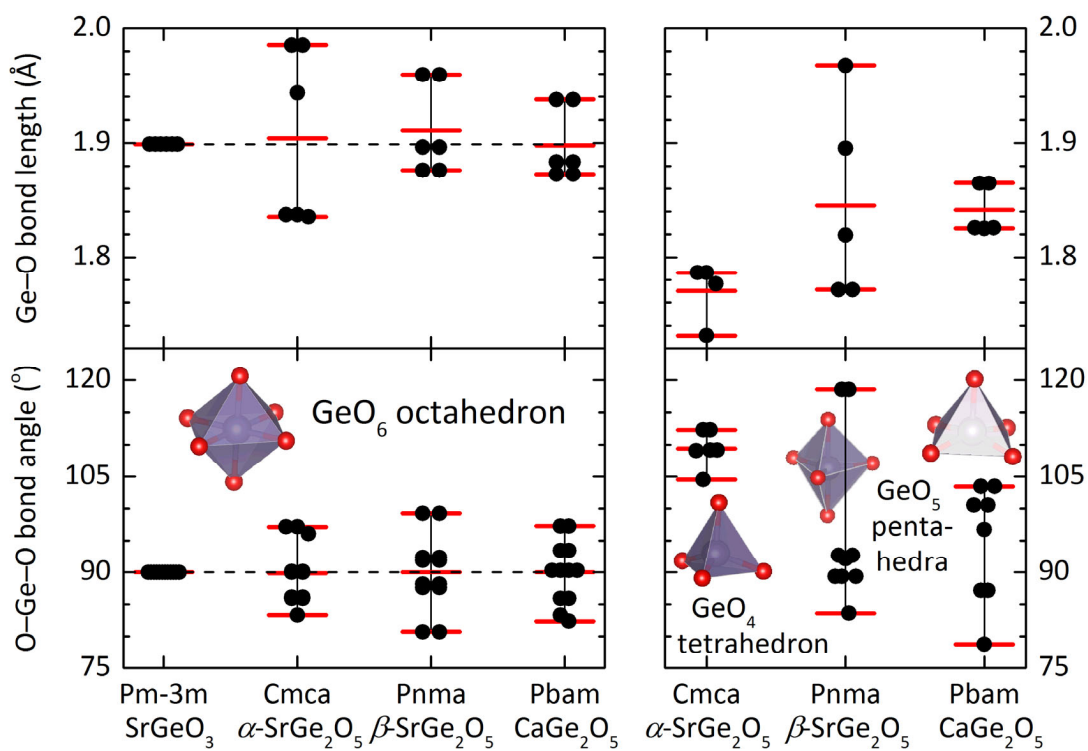

Fig. S3: Comparison of Ge-O bond lengths and O-Ge-O bond angles in the structures SrGeO$_3$ (cubic, *Pm-3m*),[3] α-SrGe$_2$O$_5$ (orthorhombic, *Cmca*),[4] β-SrGe$_2$O$_5$ (orthorhombic, *Pnma*) and CaGe$_2$O$_5$ (orthorhombic, *Pbam*)[5] for the GeO$_6$ octahedron, and GeO$_5$ and GeO$_4$ polyhedra. Red bars denote maximum, mean and minimum values, respectively.

## Calculation of the Effective Mass

Using the calculated band structure, the effective mass (Table S3) is determined from the CB curvature at the $\Gamma$ point[6]

$$m_e^* = \hbar^2 \left(\frac{d^2 E}{dk^2}\right)^{-1} \tag{1}$$

where $\hbar$ is the reduced Planck constant and $k$ is the reciprocal space vector.

Table S4: Effective mass $m_e^*$ (in units of the electron rest mass $m_e$) along high symmetry directions in the first Brillouin zone of $\beta$-SrGe$_2$O$_5$ obtained using the PBE0 hybrid functional.

| Symmetry direction | $m_e^*/m_e$ |
|---|---|
| $\Gamma$-X | 0.42 |
| $\Gamma$-Y | 0.34 |
| $\Gamma$-Z | 0.37 |
| $\Gamma$-S | 0.40 |

## Sb-doping of $\beta$-SrGe$_2$O$_5$ and IR Reflectance

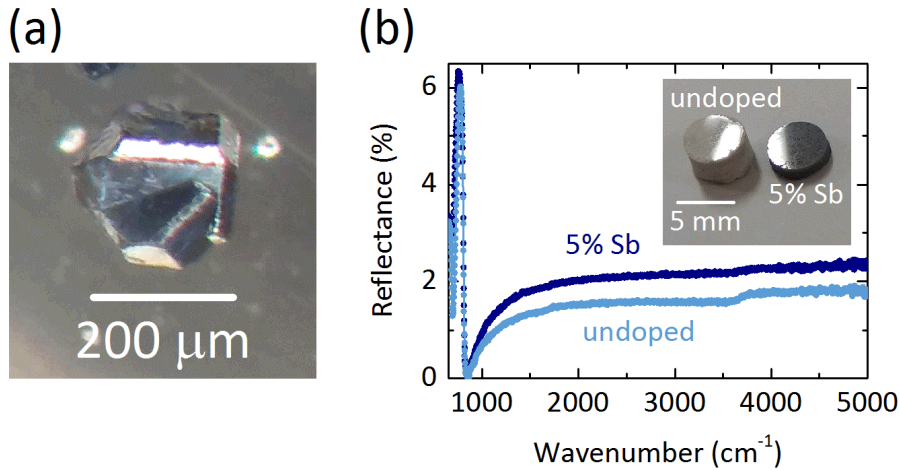

Fig. S4: (a) Optical micrograph of Sb-doped $\beta$-SrGe$_2$O$_5$ single crystal. (b) IR reflectivity spectrum of undoped $\beta$-SrGe$_2$O$_5$ and Sr(Ge$_{0.95}$Sb$_{0.05}$)$_2$O$_5$ polycrystalline pellets indicating that the reflectance edge at a wavenumber of 820 cm$^{-1}$ does not shift with Sb doping.

## The Enthalpy Equation of State

The Murnaghan equation is derived from the approximation that the bulk modulus of a material follows a linear dependence on pressure $p$ at constant temperature $T$[7]

$$-V\left(\frac{\partial p}{\partial V}\right)_T = K_0 + K_1 p \qquad (2)$$

The constants $K_0$ and $K_1$ denote the bulk modulus at zero pressure and the first derivative thereof, respectively. The Murnaghan equation is obtained after integrating Eq. (2), which yields the dependence of volume $V$ with respect to the equilibrium volume $V_0$ at zero pressure as a function of pressure $p$

$$\frac{V}{V_0} = \left(1 + \frac{K_1}{K_0} p\right)^{-\frac{1}{K_1}} \qquad (3)$$

Rearranging yields

$$p = \frac{K_0}{K_1}\left[\left(\frac{V_0}{V}\right)^{K_1} - 1\right] \qquad (4)$$

The pressure $p$ is given by the derivation of energy $E$ after volume $V$ at constant entropy $S$[8]

$$p = -\left(\frac{\partial E}{\partial V}\right)_S \qquad (5)$$

Inserting into Eq. (4), and integrating, yields the equation of state for the energy $E$ as a function of volume $V$

$$E = E_0 + \frac{K_0}{K_1} V \left[\left(\frac{V_0}{V}\right)^{K_1} \frac{1}{K_1 - 1} + 1\right] - \frac{K_0 V_0}{K_1 - 1} \qquad (6)$$

Here, $E_0$ denotes the equilibrium energy at zero pressure. The enthalpy $H$ is calculated according to[8]

$$H = E + pV \qquad (7)$$

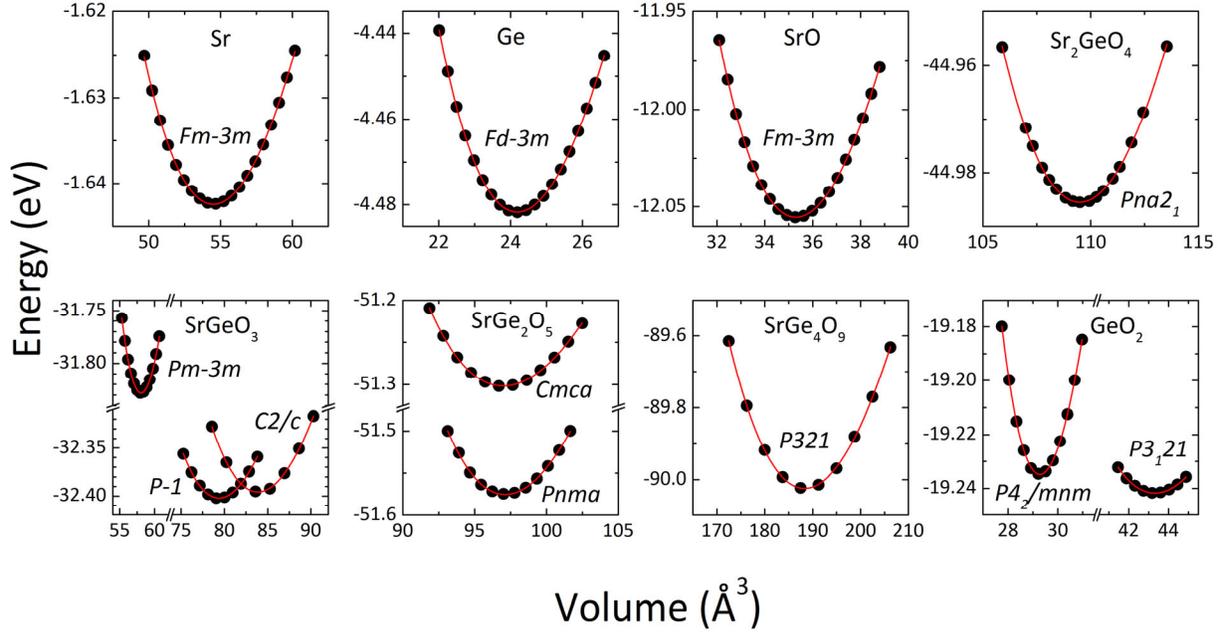

Fig. S5: PBE-calculated energy-volume dependence of elemental Sr, Ge and compounds in the SrO-GeO$_2$ system. Equilibrium energy $E_0$ and volume $V_0$ are given per unit formula. The red lines show fitting curves using the Murnaghan equation of state (Eq. (6)).

Table S5: Murnaghan equation of state (Eq. (6)) parameters for elemental Sr, Ge, and compounds in the SrO-GeO$_2$ system. Equilibrium energy $E_0$ and volume $V_0$ are given per unit formula. $K_0$ and $K_1$ denote the bulk modulus at zero pressure and the first derivative thereof, respectively.

| Material | Crystal system | Space group | $E_0$ (eV) | $V_0$ (Å$^3$) | $K_0$ (GPa) | $K_1$ - |
|---|---|---|---|---|---|---|
| Sr | cubic | Fm-3m | -1.64 | 54.51 | 11.3 | 3.47 |
| Ge | cubic | Fd-3m | -4.48 | 24.19 | 57.9 | 4.71 |
| SrO | cubic | Fm-3m | -12.06 | 35.31 | 84.4 | 4.36 |
| Sr$_2$GeO$_4$ | orthorhombic | Pna2$_1$ | -44.99 | 109.49 | 71.0 | 8.52 |
| SrGeO$_3$ | monoclinic | C2/c | -32.40 | 83.88 | 58.4 | 4.20 |
| | triclinic | P-1 | -32.40 | 79.40 | 63.4 | 4.75 |
| | cubic | Pm-3m | -31.83 | 58.06 | 157.3 | 5.10 |
| SrGe$_2$O$_5$ | orthorhombic | Pnma | -51.58 | 97.13 | 129.3 | 7.70 |
| | orthorhombic | Cmca | -51.30 | 96.99 | 91.3 | 8.70 |
| SrGe$_4$O$_9$ | trigonal | P321 | -90.02 | 188.34 | 85.5 | 3.81 |
| GeO$_2$ | trigonal | P3$_1$21 | -19.24 | 43.31 | 35.90 | 3.87 |
| | tetragonal | P4$_2$/mnm | -19.23 | 29.29 | 190.5 | 6.86 |

## Experimental Methods

We prepared $(SrO)_{1-x}(GeO_2)_x$ compounds and two-phase mixtures with the compositions $0.42 \leq x \leq 1$ by calcination of $SrCO_3$ and $GeO_2$ in stoichiometric quantities at 950–1100 °C for 12 h. The melting points and eutectic temperatures were determined by using differential thermal analysis (Bruker TG-DTA 2020SA) up to 1450 °C. A Pt crucible filled with $Al_2O_3$ powder was used as a heat transfer reference.

To grow β-$SrGe_2O_5$ single crystals, 100 mg $SrGeO_3$ (monoclinic, C2/c) powder was mixed in the 1:12 molar ratio with a $SrCl_2$/NaCl flux of the molar composition 8:2. The eutectic point in the $SrCl_2$-NaCl system lies at 544 °C and 53 mol% $SrCl_2$.[9] $SrCl_2$ and NaCl were dehydrated at 350 °C, and the powder mixture was filled into a Au capsule, pressed and inserted into a NaCl (10 wt.% $ZrO_2$) cell with a carbon heater. Using a belt-type high-pressure apparatus, a 5 GPa pressure was applied and the specimen was annealed at 1100 °C for 4 hours, then slowly cooled down to 800 °C within 18 h, at which temperature the heater was turned off and the specimen was water-cooled to room temperature. The flux was washed out with deionized water.

We performed single crystal X-ray diffraction (XRD) measurement of the β-$SrGe_2O_5$ sample at room temperature on a curved type imaging plate diffractometer (Rigaku R-AXIS RAPID II) with graphite monochromatic Mo-$K_α$ radiation. The intensity data was collected by the Rapid Auto program (Rigaku). The initial structural model (orthorhombic, Pmna), named β-$SrGe_2O_5$, was suggested based on the unit cell dimensions and the reflection data. Internal atomic coordinates and anisotropic displacement parameters for all atoms were then refined according to the full-matrix least-squares method using SHELXL.[10] The Bragg-Brentano powder XRD pattern of ground single crystals was recorded using a Bruker RINT2000 diffractometer with a Cu $K_α$ source.

Undoped β-$SrGe_2O_5$ and Sb-doped β-$Sr(Ge_{0.95}Sb_{0.05})_2O_5$ polycrystalline pellets were prepared from stoichiometric quantities of $SrCO_3$, $GeO_2$ and $Sb_2O_5$, ground and calcined at 950 °C for 12 h. The powder mixture was pressed into a Cu capsule, then sintered at 1000 °C for 2 h at 4 GPa pressure. For optical measurements, the pellet surface was mirror-polished by finishing with 0.3 µm fine-grained foil. The Fourier-transform infrared (FTIR) reflectance spectrum (Bruker Vertex 70v) of the polished β-$SrGe_2O_5$ and β-$Sr(Ge_{0.95}Sb_{0.05})_2O_5$ pellets were measured from 500-5000 $cm^{-1}$ wavenumber using a Au mirror as a 100% reflectance standard.

## Computational Methods

Hybrid DFT calculations using the PBE0 functional[11] were performed to obtain the β-SrGe$_2$O$_5$ (orthorhombic, *Pnma*) ground state crystal structure, with the ionic forces converged to less than 0.01 eV/Å. We used the projector augmented wave (PAW) method as implemented in the Vienna Ab initio Simulation Package (VASP).[12,13] The electronic band structure was calculated employing maximally localized Wannier functions,[14] a Γ-centered 4 x 4 x 3 *k*-mesh and a plane wave cutoff energy of 500 eV. We employed the PBE0 functional because it gives a reasonable band gap for related compounds, SrGeO$_3$ (cubic, *Pm-3m*, $E_g^{calc}$ = 2.74 eV, $E_g^{exp}$ = 2.8 eV) and GeO$_2$ (tetragonal, *P4$_2$/mnm*, $E_g^{calc}$ = 4.55 eV, $E_g^{exp}$ = 4.68 eV[15]). The frequency-dependent real and imaginary dielectric tensor elements $\varepsilon_x$, $\varepsilon_y$ and $\varepsilon_z$ were calculated using density functional perturbation theory.[16]

Employing DFT, energy calculations of all (meta)stable compounds including the high-pressure polymorphs in the SrO-GeO$_2$ system were performed for a fixed set of volumes around the equilibrium volume. In each calculation, the unit cell shape was relaxed at the given constant volume. We used variable-cell structure relaxations using the generalized gradient approximation (GGA) and the Perdew-Burke-Ernzerhof (PBE) functional.[17] A maximum *k*-point spacing of 0.3 Å$^{-1}$ was used.